\input{aipcheck}

\documentclass[
    ,final            % use final for the camera ready runs
  ]
  {aipproc}

\layoutstyle{6x9}

\begin{document}

\title{White dwarfs with jets as non-relativistic 
        analogues of quasars and microquasars?}
        
 \classification{}
 \keywords      {}

\author{R. Zamanov}{
  address={Astrophysics Research Institute, Liverpool John Moores University, UK}
}

\author{M.F.Bode}{
  address={Astrophysics Research Institute, Liverpool John Moores University, UK}
 % ,altaddress={<author1 address>}
}

\author{P.Marziani}{
  address={Osservatorio Astronomico di Padova, INAF,  Padova, Italy}
 % ,altaddress={<author1 address>} % additional visiting address
}
\author{R.J. Davis}{
  address={Jodrell Bank Observatory, University of Manchester, UK }
 % ,altaddress={<author1 address>} % additional visiting address
}
\author{ S.P.S. Eyres}{
  address={Centre for Astrophysics, University of Central Lancashire, UK}
 % ,altaddress={<author1 address>} % additional visiting address
}
\author{ A. Gomboc}{
 address={Astrophysics Research Institute, Liverpool John Moores University, UK}
 % address={<common address for author2 and author3>}
 % ,altaddress={<author1 address>} % additional visiting address
}
\author{J. Porter }{
 address={Astrophysics Research Institute, Liverpool John Moores University, UK}
%  address={<common address for author2 and author3>}
%  ,altaddress={<author1 address>} % additional visiting address
}
\author{A. Skopal }{
  address={Astronomical Institute, Slovak Academy of Sciences, Slovakia}
 % ,altaddress={<author1 address>} % additional visiting address
}

\begin{abstract}
We explore the similarities between accreting white dwarfs  (CH Cyg and MWC 560) and  the much 
more energetic jet sources - quasars and microquasars.  To-date we have identified several common  
attributes:  
	(1) they exhibit collimated outflows (jets);
	(2) the jets are precessing;
	(3) these two symbiotic stars exhibit quasar-like emission line spectra;
	(4) there is a disk-jet connection like that observed in microquasars.
Additionally they may have a similar energy source (extraction of rotational energy from the 
accreting object). Study of the low energy analogues could have important implications for our 
understanding of their higher energy cousins.
\end{abstract}

\maketitle

%%%%%%%%%%%%%%%%%%%%%%%%%%%%%%%%%%%%%%%%%%%%
%% MAINMATTER
%%%%%%%%%%%%%%%%%%%%%%%%%%%%%%%%%%%%%%%%%%%%
\section{Emission line similarities}
As illustrated in Zamanov \& Marziani (2002), 
there are striking similarities between the optical spectra of 
Active Galactic Nuclei (AGNs) and two accreting white dwarfs. Almost every emission line visible 
in the  AGN spectrum of I Zw 1 shows a corresponding  feature in the spectra of CH Cyg and MWC 560. 
The similarity between the UV spectrum of CH Cyg and  I Zw 1 is demonstrated in Fig.1.

In AGN, hydrogen and  FeII emission lines  are emitted from the so-called Broad Line Region. 
This region is thought to lie within $<$1 pc of the central black hole. Its structure is still poorly 
understood. The clear similarity between the emission lines  
suggests  that  we are observing a scaled down version of the quasar 
Broad Line Region in galactic objects like MWC560 and CH Cyg 
(see also Zamanov and Marziani, 2002).
\begin{figure}
 \includegraphics[width=15.0cm]{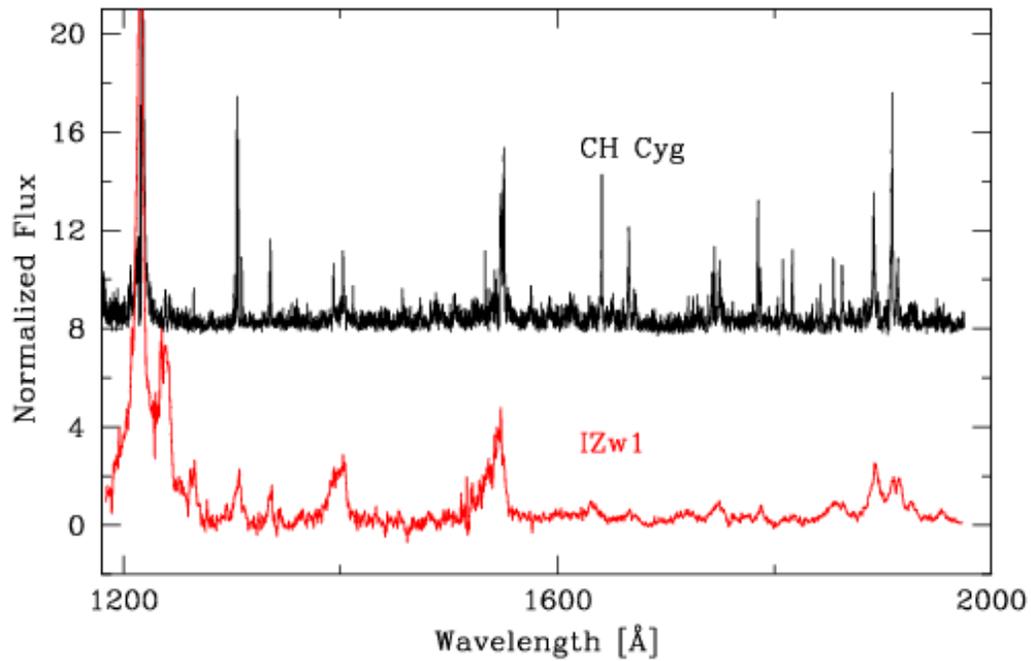}
 \caption{The UV spectra of symbiotic star CH Cyg and Narrow Line Seyfert 1
 galaxy I~Zw~1. Clear similarity in the emission lines is visible.}
\end{figure}
% \begin{figure}
% \includegraphics[width=5cm,height=5cm,scale=1.0,clip=true, draft=true]{fig1uv.eps}
% \includegraphics[width=15.0cm]{fig1uv.ps}
% \caption{Picture to fixed height}
% \end{figure}

\section{Disk-Jet connection}

\begin{figure}
 \includegraphics[height=.3\textheight]{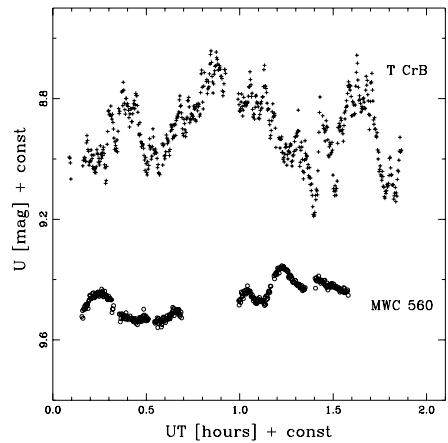}
 \caption{The flickering behaviour of MWC~560 and T~CrB. }
\end{figure}
Comparison of the flickering behaviour of T CrB (February 28, 1995) and MWC 560 (March 05, 1990)
is shown on Fig. 2. In the case of MWC 560, the short term variability 
(on a time scale of minutes) is missing. 
Only smooth, hour-timescale variations are present. 
This indicates disruption of the inner part of the accretion disk during the time of jet ejection. 
 A disruption of the inner disk (disk-jet) connection is also observed 
 in CH Cyg (Sokoloski \& Kenyon 2003). 
 The behaviour is closely analogous to that of the microquasar GRS1915+105. 
 This supports the view that there may be a common mechanism for jets in quasars, 
 microquasars and symbiotic stars (see also Livio, Pringle \& King 2003). 

\section{Precessing jets}
The best known precessing jets in astrophysics 
are probably those of SS 433. In recent years precessing jets have 
been identified in two symbiotic stars: CH Cyg (on the basis of radio imaging by Crocker et al. 2002)
and MWC 560 (from optical spectroscopy by Iijima 2002). In both cases the model of the jets of the 
microquasar  SS 433 has been adopted to fit the evolution of the morphology of the outflows 
(using velocities appropriate for  white dwarfs).

%\begin{figure*}
%\centering
%\includegraphics[width=1cm]{f1.eps}
%\caption{VLA 5 GHz images of CH Cyg from 1986-2000 compared with the simple ballistic model 
%(dashed lines). Each box is ~4 arcsec on a side (from Crocker et al. 2002). 
%Variation of the position angle of the extended emission in the central region of CH Cyg between 
%1985 and 2000, along with a curve representing the precessing jet model, adapted from that for 
%the high-velocity jets of SS 433}
%\label{cool}
%\end{figure*}
\begin{figure}
 \includegraphics[height=.3\textheight]{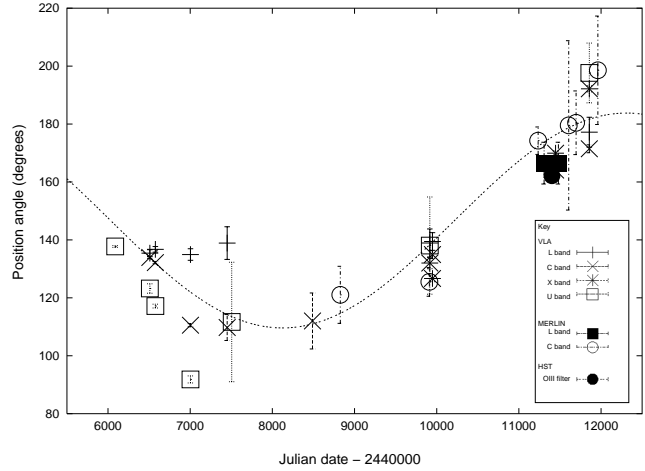}
 \caption{Variation of the position angle of the extended emission in the central region 
 of CH Cyg between 1985 and 2000, along with a curve representing the precessing jet model, 
 adapted from that for the high-velocity jets of SS 433.}
\end{figure}

%\section{Disk jet connection}
%\begin{figure*}
%\centering
%\includegraphics[width=1cm]{f1.eps}
%\caption{Fig.7. Comparison of the flickering behaviour of T CrB (February 28, 1995) and MWC 560 
%(March 05, 1990). In the case of MWC 560, the short term variability (on a time scale of minutes) 
%is missing. Only smooth, hour-timescale variations are present. This indicates disruption of the 
%inner part of the accretion disk during the time of jet ejection.  A disk-jet connection is also 
%observed in CH Cyg (Sokoloski and Kenyon 2003). The behaviour is closely analogous to that of the 
%microquasar GRS1915+105. This supports the view that there may be a common mechanism for jets in 
%quasars, microquasars and symbiotic stars (see also Livio, Pringle and King 2003). 
%}
%\label{cool}
% \end{figure*}

\section{Energy Source of Jets}
It is worth noting that interacting binaries, where a white dwarf accretes material from the wind 
of a red giant (usually classified as symbiotic stars), are strongly variable objects. We show above 
that the spectra of CH Cyg and MWC 560 are similar to low-redshift quasars around the times when jet 
activity is detected (CH Cyg - July 1984, MWC 560 - November 1990).

Jets are detected in systems quite different from those harboring black holes (for a review see 
Livio, 2001): young stellar objects (velocity v$\sim$200 km s$^{-1}$), planetary nebulae 
(v$\sim$200-1000 km s-1),
supersoft X-ray sources (v$\sim$1000 km s$^{-1}$). 
The jet velocities observed in the accreting white dwarfs 
(we call them {\it ``nanoquasars''}) are  $\sim$1000 km s$^{-1}$ in CH Cyg 
(Taylor et al. 1986) and 1000-6000 km s$^{-1}$  in MWC 560 (Tomov et al. 1992). 
They are consistent with an overall picture in which the jet 
velocity is of the same order as the escape velocity from the accretor (Livio 2001).

The luminosities of MWC 560 and CH Cyg are considerably less than the Eddington limit. The mass 
accretion rate is about 
M$_{acc} \sim $0.05 M$_{Edd}$. At such mass accretion rates the most probable jet energy 
source  involves extraction of rotational energy from the compact object. In the case of 
nano-quasars the extraction is probably occurring via the propeller action of a magnetic white 
dwarf (Mikolajewski et al. 1996). The most probable source of jet formation in quasars is the 
extraction of energy and angular momentum via the Blandford and Znajek (1977) mechanism. In this 
sense the jets in the ``nanoquasars''  probably represent a low energy (non-relativistic) analogue 
of the jets in quasars and microquasars. They involve a  similar energy source - the extraction of 
rotational energy from the central compact object.

\section{Conclusions}
We think it is appropriate to call the two accreting white dwarfs discussed here "nanoquasars" 
because they represent the very low energy analogue of quasars and microquasars. The name is 
chosen by analogy with the quasar and microquasar denominations, and also because 
$\nu \alpha \nu  o  \varsigma $
(ancient greek) = nano (ital.) = dwarf(engl.).

We suggest that the "nanoquasars" could be  an important link in our understanding of  a broad 
range of  accreting sources. They could help us to create a unified picture of accreting objects 
from cataclysmic variables and stellar-mass black holes up to the most powerful quasars. 

\vskip 0.4cm 
{\bf References}  \\
Blandford R. \& Znajek R., 1977, MNRAS 179, 433  \\
Crocker et al., 2002, MNRAS 335, 1100  \\
Iijima, T., 2002, A\&A 391, 617 \\
Livio M., 2001, ASP Conf. Ser., v.224, p.225  \\
Livio M., Pringle J.E., King A.R., 2003, ApJ 593, 184 \\
Mikolajewski, M., Milkolajewska, J., Tomov, T. 1996, IAUS 165, 451 \\
Sokoloski J. \& Kenyon S., 2003, ApJ 584, 1021    \\
Taylor  A.R., Seaquist E.R., Mattei J.A., 1986, Nature 319, 38  \\
Tomov T., Zamanov R., Kolev D., et al. 1992, MNRAS 258, 23 \\
Zamanov, R., Marziani, P., 2002, ApJ 571, L77 \\

\end{document}